\documentclass[aps,prl,twocolumn,groupedaddress]{revtex4}
\usepackage{graphicx}

\begin{document}


\title{Controlling spin current in a trapped Fermi gas}


\author{X.~Du, Y.~Zhang, J.~Petricka, and J.~E.~Thomas}
\email[jet@phy.duke.edu]{}
\affiliation{Duke University, Department of Physics, Durham, North
Carolina, 27708, USA}



\date{\today}

\begin{abstract}

We study fundamental features of spin current in a very weakly
interacting Fermi gas of $^6$Li. By creating a spin current and then
reversing its flow, we demonstrate control of the spin current. 
This reversal is predicted by a spin vector evolution equation in energy representation, which shows
how the spin and energy of individual atoms become correlated in the nearly undamped regime of the experiments.
The theory provides a simple physical description of  the spin current and explains both the large amplitude and the slow temporal evolution of the data. Our results have applications
in studying and controlling fundamental spin interactions and spin
currents in ultra-cold gases.

\end{abstract}

\pacs{313.43}

\maketitle

Spin dynamics and spin currents have been extensively studied in
condensed matter physics~\cite{RMP2004}. Active manipulation of the
electron spin can be used for data processing and
storage~\cite{AmScientist2001}. For example, a spin current can be
used to  excite or reverse the magnetization of a
nanomagnet~\cite{Duine2008,Slonczewski1996, Berger1996}. In
ultracold atomic physics, spin-current-related phenomena have been
observed both in a Bose gas~\cite{Lewandowski2002} and in a Fermi
gas~\cite{Du2008}. We report on the origin and control of spin
current in a weakly interacting Fermi gas.

An optically trapped Fermi gas of $^6$Li is a rich system  in which
the strength of interactions between atoms can be controlled by
applying a variable bias magnetic field tuned near a Feshbach
resonance~\cite{Feshbach1,Feshbach2}. Close to resonance,  the Fermi
gas exhibits strong interactions~\cite{O'Hara2002}, which have been
widely studied~\cite{RMP2008}. This regime offers unprecedented
opportunities to test nonperturbative quantum many-body theories. In
contrast, little investigation has been done for a weakly
interacting Fermi gas, where the s-wave scattering length, a$_{12}$,
for atoms in opposite spin states can be tuned smoothly from small
and positive to small and negative. In this regime, spin segregation
is observed, where atoms of one spin move outward in the trap, while
atoms with the opposite spin move inward. Previous observations of
spin segregation~\cite{Du2008} have shed new light on the study of
this regime.

An overdamped spin wave theory~\cite{Oktel2002, Fuchs2002,
Williams2002} has been used to explain the spin segregation observed
in a Bose gas of $^{87}$Rb confined in a magnetic
trap~\cite{Lewandowski2002}. In those experiments, the collision
rate between the atoms was large compared to the axial trap
frequency and sufficient to ensure a thermal momentum distribution.
The predictions are in good agreement with the measurements for the
Bose gas. In contrast, the corresponding theory of overdamped spin
waves  for a Fermi gas disagrees with the experiments by two orders
of magnitude in amplitude, and predicts an oscillation of the
density profile, which is not observed~\cite{Du2008}.

In the Fermi gas experiments with $^6$Li, collisions between atoms
in the same state are prohibited due to the Pauli principle, and the
scattering length between atoms in opposite spin states is
magnetically tuned to be very small. In this case, velocity changing
collisions between atoms in different states occur at a rate of only
$\sim$0.3 Hz, small compared to both the axial trap frequency and
the spin segregation rate. We explain the observed spin segregation
in this regime by a nearly undamped spin wave, in which the spin
vector of each atom is correlated with its energy.

\begin{figure}
\begin{center}\
\includegraphics[width=2.5in]{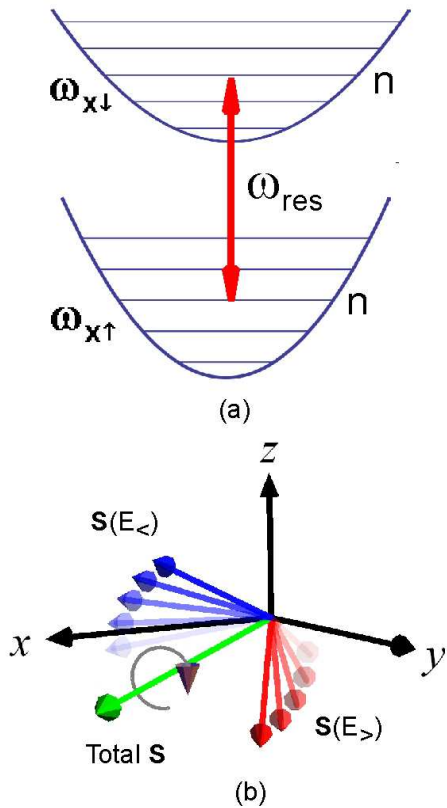}
\end{center}
\caption{Spin-wave formation. {\bf (a)} For an rf transition between
harmonic oscillator-spin states $\mid n,\uparrow\rangle$ and $\mid
n,\downarrow\rangle$, the resonance frequency $\omega_{res}$
(denoted by the red arrow) increases with $n$, due to the difference
in the harmonic oscillator frequencies for the two spin states. An
rf pulse initially creates x-polarized spins (in the rotating
frame). {\bf (b)} The spin vector for atoms of high energy $E_>$
precesses more than for atoms of low energy $E_<$. Binary collisions
then cause the spin vectors to rotate about the total spin vector
$\mathbf{S}$, producing a $z$-polarized spin wave.
\label{fig:spin-interaction}}
\end{figure}

To understand the origin of the spin-segregation and corresponding
current, consider the evolution of the spin vectors of atoms
vibrating almost freely along the axial (long) direction of a
cigar-shaped optical trap. Atoms in states $\mid \uparrow\rangle$
and $\mid \downarrow\rangle$ oscillate in the trap with frequencies
$\omega_{x\uparrow}$ and $\omega_{x\downarrow}$, respectively. For
the trap, $\overline{\omega}_x\equiv
(\omega_{x\uparrow}+\omega_{x\downarrow})/2=2\pi\times 145$ Hz. As
the magnetic moments of the two spin states are not identical, the
finite curvature of the bias magnetic field causes a small
difference in the axial confining potentials and hence in the
oscillation frequencies for the two spin states,
$\delta\omega_x\equiv\omega_{x\downarrow}-\omega_{x\uparrow}=-2\pi\times2.5$
mHz (the corresponding difference in the transverse oscillation
frequencies in the optical trap is negligible, due to the tight
transverse confinement). The small difference in the axial
frequencies correlates the precession rate $\Omega (E)$ of an atomic
spin vector in the $x-y$ plane with the energy of the atom. This
along with binary collisions causes a spin wave.

Fig.~\ref{fig:spin-interaction}(a) shows that the resonance
frequency $\omega_{res}$ for a radio-frequency (rf) transition
between states $|n,\uparrow\rangle$ and $|n,\downarrow\rangle$ is
shifted by $(n+1/2)(\omega_{x\downarrow}-\omega_{x\uparrow})$, where
$n$ is the harmonic oscillator quantum number for the axial
direction, which does not change in the transition. The axial energy
$E$ of the atom determines $n+1/2=E/\hbar\overline{\omega}_x$. As
the collision rate for the Fermi gas of $^6$Li in the experiment is
very low compared to the axial trap frequency, collisions do not
significantly change the energy of each atom over the time scale of
the segregation, and the precession rate can be written as
\begin{equation}
\Omega (E)=-(\delta\omega_x/\overline{\omega}_x)E/\hbar.
\label{eq:shift}
\end{equation}
As a result of this energy-dependent precession rate, the magnitude
of the precession angle of the spin vector in the $x-y$ plane is
larger for atoms with high energy than for  atoms with low energy.

When two coherently-prepared atoms collide, the energy-dependent
precession angle then leads to a correlation between the
$z-$component of the spin vector and the energy.  The collisional
interaction results in a rotation of each atom's spin vector about
the total spin vector, which is conserved. The sense of the rotation
is determined by the sign of the scattering length $a_{12}$, and the
relative angle of the spin vectors. As both atoms have spins in the
\emph{x-y} plane, the rotation of each spin about the total spin in
the \emph{x-y} plane produces spin components out of the \emph{x-y}
plane, as shown in Fig.~\ref{fig:spin-interaction}(b). For a
positive (negative) scattering length, atoms with higher energy
$E_>$ will accumulate a negative  (positive) $z-$component, while
atoms with lower energy $E_<$ will accumulate a positive  (negative)
$z-$component. This process correlates the $z-$component of the spin
with the energy, i.e., $S_z(E)$.

The spin density vector in coordinate space is then determined by
the axial harmonic oscillator wavefunctions $\phi_E(x)$ as
$\mathbf{S}(x,t)=\int dE\, \mathbf{S}(E,t)|\phi_E(x)|^2$, where we
assume that there is no coherence between different energy states.
The energy-dependent spin vector $\mathbf{S}(E,t)$ is determined
from the Heisenberg equations
$\dot{\hat{\mathbf{S}}}(E,t)=(i/\hbar)[\hat{H},\hat{\mathbf{S}}]$,
where the components of $\hat{\mathbf{S}}(E,t)$ are written in terms
of creation and annihilation operators in energy representation.

In a one-dimensional approximation, the Hamiltonian operator for a
Fermi gas in the optical trap is
\begin{equation}\label{H}
\hat{H}=\hat{H}_0+\hat{H}_{int}.
\end{equation}
In a frame rotating at the unshifted hyperfine transition frequency,
\begin{equation}\label{H0}
  \hat{H}_0 =
  \sum_{E}E\left(\hat{N}_\uparrow(E)+\hat{N}_\downarrow(E)\right)+
    \sum_{E}\hbar\Omega(E)\hat{S}_z(E) ,
\end{equation}
where  $\hat{N}_{\uparrow,\,\downarrow}(E)$ are the number operators
for each state and
$\hat{S}_z(E)=[\hat{N}_{\uparrow}(E)-\hat{N}_{\downarrow}(E)]/2$.

Collisions produce a contact interaction between opposite spin
states. Averaging over the transverse coordinates ($z,y$), the
collision operator takes the form
\begin{equation}
 \hat{H}_{int} = \frac{4\pi\hbar^2 a_{12}}{m}\frac{1}{2\pi\sigma^2_{\rho}}
 \int dx\,\hat{\psi}_\uparrow^\dagger(x)\hat{\psi}_\downarrow^\dagger(x)
 \hat{\psi}_\downarrow(x)\hat{\psi}_\uparrow(x).
 \label{eq:Hint}
\end{equation}
Here, $m$ is the atomic mass and $\sigma_\rho$ is radial $1/e$ width
for a fit of a Gaussian distribution of the trapped cloud.
Eq.~\ref{eq:Hint} can be written in energy representation using
$\hat{\psi}_{\uparrow,\,\downarrow}(x)=\int
dE\,\hat{a}_{\uparrow,\,\downarrow}(E)\,\phi_E(x)$, where
$\phi_E(x)$ is the axial harmonic oscillator wavefunction.

Using $\mathbf{S}(E,t)=\langle\hat{\mathbf{S}}(E,t)\rangle$, we
obtain the evolution equations
\begin{eqnarray}\label{eq:spinvsE}
   \nonumber \frac{\partial \mathbf{S}(E,t)}{\partial t} &=&
   \mathbf{\Omega}(E)\times\mathbf{S}(E,t)
    \\
    &+&\int dE'\,g(E',E)\,\mathbf{S(E',t)}\times\mathbf{S(E,t)},
\end{eqnarray}
where
\begin{eqnarray}\label{eq:g}
    g(E',E)&=&-\frac{4\pi\hbar a_{12}}{m}\frac{1}{\pi\sigma^2_{\rho}}\left(\frac{m\omega_x^2}{2\pi^4\,E_{min}}\right)^{1/2}\nonumber\\
    & &
    \int_{-\frac{\pi}{2}}^{\frac{\pi}{2}}\frac{d\theta}{\left(\frac{|E-E'|}{E_{min}}+\cos^2\theta\right)^{1/2}}.
\end{eqnarray}
Here $E_{min}=\min(E,E')$ and we have assumed a WKB
approximation~\cite{Sakurai} for $\phi_E(x)$ between the classical
turning points, since the energies are in the classical regime.

\begin{figure}
\begin{center}\
\includegraphics[width=3.2in]{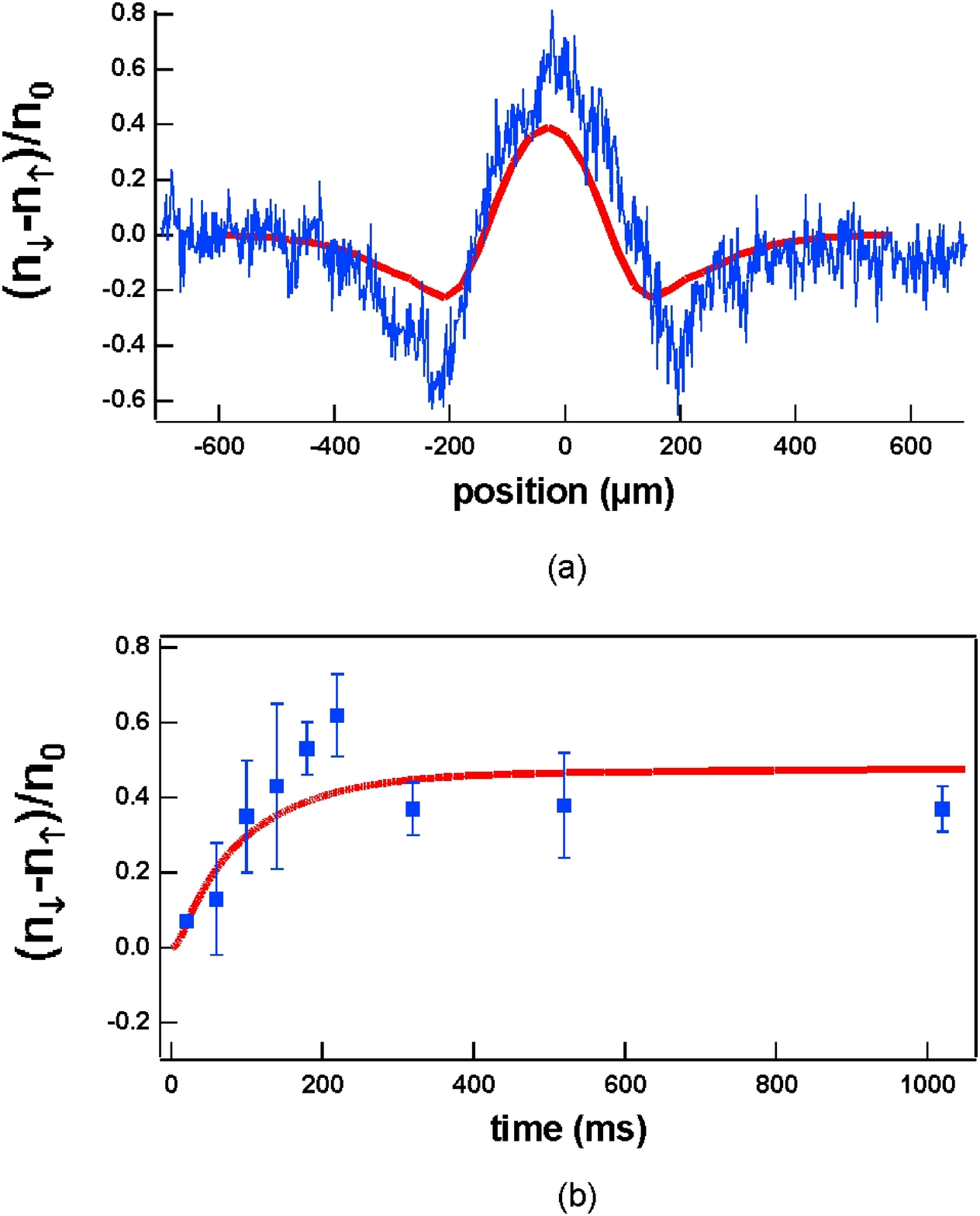}
\end{center}
\caption{{\bf (a)} Difference between the spin-down and spin-up
densities $n_\downarrow-n_\uparrow$ at 220 ms in units of
$n_0=(n_{\downarrow 0}+n_{\uparrow 0})/2$. Here $n_{i0}$ is the
initial spin density for each state at the trap center before spin
segregation occurs. Data (blue line); Theory (red curve).  The data
is taken for $a_{12}=-4.5 a_0$; {\bf (b)} $n_\downarrow-n_\uparrow$
at the trap center versus time for $a_{12}=-4.5 a_0$. Data (blue
dots); Theory (red curve). \label{fig:spin-segregation}}
\end{figure}

Eq.~\ref{eq:spinvsE} is a primary result of this paper. The first
term describes the energy-dependent spin precession, while the
second term describes the rotation of the spin vector arising from
binary collisions. The initial spin vector is  $S_x(E,t=0)=
N\exp(-E/K_BT)/(2k_BT)$, where $N$ is the total number of atoms and
the factor $1/2$ arises from the definition of the spin vectors.
Fig.~\ref{fig:spin-segregation} shows that the predictions obtained
by numerical integration of Eq.~\ref{eq:spinvsE} are in good
agreement with the data and trap parameters of Ref.~\cite{Du2008}.

Our description of the spin vector evolution  in energy
representation can be compared to two recent theories based on a
collisionless Boltzmann equation~\cite{Laloe,Mueller}. These
approaches provide a phase space description in one dimension for a
weakly interacting two-component Fermi gas. Their results are in
very good agreement with the predictions of Eq.~\ref{eq:spinvsE},
both in amplitude and temporal evolution.

\begin{figure}
\begin{center}\
\includegraphics[width=3.5in]{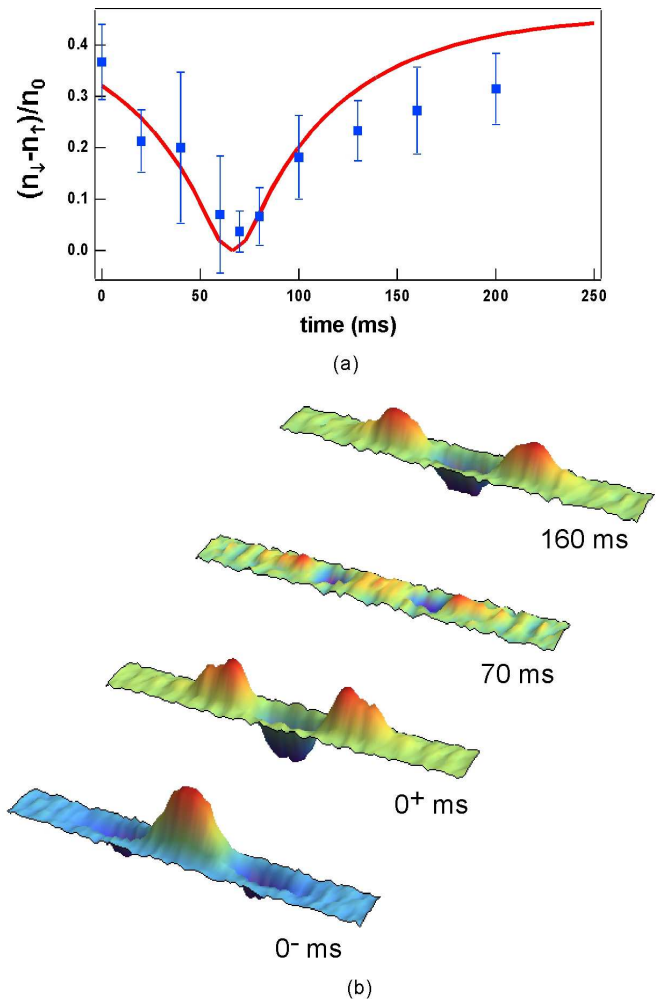}
\end{center}
\caption{Reversal of the spin current in an ultracold Fermi gas.
{\bf (a)} Difference in the spin densities $n_\downarrow-n_\uparrow$
at the trap center versus time, showing return to $0$ at $t\simeq 70$ ms, after an rf
$\pi$ pulse is applied and the sign of scattering length is
inverted at $t\simeq 0$. Prediction (red line); Data (blue dots). {\bf (b)} Images
of the spin waves corresponding to the time in {\bf (a)}. Each image
is the z-component of the spin density as a function of position $x$
along the axial direction of the trap. 0$^-$ ms corresponds to time
just before the rf $\pi$ pulse and 0$^+$ ms corresponds to time just
after the rf $\pi$ pulse and reversal of the scattering length.
\label{fig:re-phasing}}
\end{figure}

The spin-energy correlation description is particularly useful when
the evolution is nearly Hamiltonian, where it provides a very simple
physical picture for a weakly interacting spin system.
Eq.~\ref{eq:spinvsE} makes it apparent that the spin current can be
reversed, causing the spatial distribution to return to the
unsegregated state. Two steps are required: A $\pi$ pulse is applied
to reverse the sign of $S_z$ and either $S_y$ or $S_x$; the sign of
the scattering length is inverted by sweeping the bias magnetic
field through the zero crossing.  Then the cloud will start to
merge. The quantitative predictions of our numerical simulations are
shown in Fig.~\ref{fig:re-phasing}(a). Our experiments confirm this
prediction. Immediately after the rf $\pi$ pulse and inversion of
the sign of the scattering length, the difference in the spin-up and
spin-down densities at the cloud center is 30\% of the average
density. As the cloud merges,  this difference decreases to zero in
$\sim70$ ms, in contrast to the case without current reversal, where
spin segregation persists for a few seconds. Then the density
difference continues to evolve and returns to 30$\%$.

To perform these experiments, we prepare a spin segregated sample of
$^6$Li Fermi gas at $a_{12}=8.1 a_0$~\cite{Grimm2005} (the bias
magnetic field $B=529.8$ G is calibrated with rf spectroscopic
techniques).  A sample of $^6$Li atoms in a 50-50 mixture of the two
lowest hyperfine states is loaded into a CO$_2$ laser trap with a
bias magnetic field of 840 G, where the two states are strongly
interacting. Evaporative cooling is performed to lower the
temperature of the sample~\cite{O'Hara2002}. The magnetic field is
then increased in 0.8 seconds to  a weakly interacting regime at
1200 Gauss where an on-resonance optical pulse of 40 $\mu$s is
applied to remove atoms of one state, while leaving atoms in the
other state. With a single state present, the magnetic field is
lowered in 0.8 seconds to 529.8 G. Then a 40 ms rf pulse (center
frequency 75.613 MHz and sweep range 35 kHz) is applied on the
$\mid\uparrow\rangle-\mid\downarrow\rangle$ transition to create a
50-50 coherent superposition of the two spin states. Note that, as
the frequency passes through resonance, coherence is created on a
time scale of a few milliseconds, short compared to the time for the
total sweep and for spin segregation to occur. At the final optical
trap depth, the measured trap oscillation frequency in the
transverse directions is $\omega_\bot=2\pi\times 3900$
 Hz, while the axial frequency is $\omega_x=2\pi\times 120$ Hz. The total number of atoms
 is $N\simeq 4.0\times 10^5$. The
corresponding Fermi temperature is $T_F\simeq6$ $\mu$K. The sample
temperature is $T\simeq 33$ $\mu$K. The peak atomic density at $T$
is $6\times10^{11}$/cm$^3$. The axial and radial 1/e widths for a
fit of a gaussian distribution to the initial density profile of the
sample are $\simeq 400 \,\mu$m and $\simeq 12 \,\mu$m, respectively.

At 40 ms after the first rf pulse, when difference in the spin-up
and spin-down densities is 30\% of the average density, we change
the bias magnetic field from 529.8 G ($a_{12}= 8.1 a_0$) to 525.2 G
($a_{12}=-8.1 a_0$) in 5 ms. Then  we apply an rf $\pi$ pulse
(duration 40 ms, center frequency 75.596 MHz and sweep range 4 kHz),
which flips the spins, as shown in the $0^-$ and $0^+$ images of
Fig.~\ref{fig:re-phasing}(b). Finally, we take absorption images of
atoms in both states (in separate experimental cycles) at various
times after the rf $\pi$ pulse,
Fig.~\ref{fig:re-phasing}~\cite{note}. The entire time sequence is
done in random order and repeated 6 times. The error bars are
statistical and arise from run-to-run variations in the atom number,
magnetic field and excitation frequency.

As a further test of this idea, we applied either the scattering
length sign change or the rf $\pi$ pulse, but not both, and looked
for merging followed by segregation. We found no reversal of spin
segregation. This verifies that both operations are required to
observe the reversal of spin segregation, as predicted.

In an additional experiment, we increased the length of the initial
spin segregation time, to determine the longest time scale $\tau$ over which the
spin segregation is reversible.  We expect that $\tau$ must be significantly smaller than the velocity changing collision time ($\sim3$ sec) for reversal to occur and find $\tau = 200$ms.

Our experiments suggest that broad manipulation of the spin dynamics
and creation and study of non-equilibrium systems in arbitrary spin
mixtures is possible, by using general rf pulse sequences and by
temporally varying the scattering length and the spatial dependence
of the magnetic field.

This research is supported by the Physics Divisions of the Army
Research Office and the National Science Foundation, and the
Chemical Sciences, Geosciences and Biosciences Division of  the
Office of Basic Energy Sciences, Office of Science, U.S. Department
of Energy. We thank Paul Julienne for providing the magnetic field
dependence of the scattering length near the zero crossing and Le
Luo and Bason Clancy for help during the initial stages of these
experiments. Finally, we are indebted to Franck Lalo\"{e} for
stimulating discussion and for pointing out an error in  Eq. 6 of
the original manuscript.


\begin{references}
\bibitem{note} The numerical simulation shows the merging time,
i.e., time for segregation to disappear, is equal to the waiting
time for segregation before the second rf pulse. The waiting time is
estimated as 65 ms, including 40 ms (time period between end of the
first rf pulse and start of magnetic field switch), 20 ms (half of
the first rf pulse duration), and 5 ms (magnetic field switch time).
\end{references}
 \end{document}